\documentstyle[pre,aps]{revtex}
\begin{document} 
\newcommand{\beqa}{\begin{eqnarray}}
\newcommand{\eeqa}{\end{eqnarray}}
\newcommand{\beq}{\begin{equation}}
\newcommand{\eeq}{\end{equation}}

\title{Nonlinear Dynamics of Aeolian Sand Ripples} 
\author{Leonid Prigozhin}
\date{\today} 
\address{Center for Energy and Environmental Physics, Blaustein Institute 
for Desert Research,\\Ben Gurion University of the Negev,
Sede Boqer Campus, 84990 Israel}

\maketitle

\begin{abstract} We study the initial instability of flat sand surface
and further nonlinear dynamics of wind ripples.  The proposed continuous
model of ripple formation allowed us to simulate the development of a
typical asymmetric ripple shape and the evolution of sand ripple
pattern. We suggest that this evolution occurs via ripple merger
preceded by several soliton-like interaction of ripples. 
\end{abstract} 

\pacs{PACS numbers: 81.05.Rm, 47.54.+r, 92.60.Gn}

\section{Introduction} 
Aeolian (i.e. wind driven) sand ripples form nice regular patterns on
coastal beaches and desert floors and indicate an instability
of flat sand surface under the wind-induced transport and rearrangement of
loosely packed sand grains. Following fundamental work by Bagnold 
\cite{Bagnold} 
more than half a century ago, formation of sand ripples has been studied by 
many researchers (see review \cite{And90} and the references therein). 
Significant progress in understanding the nature of this phenomenon
has been achieved. Nonetheless, major questions remain open; these 
involve the most interesting part of ripple formation, the nonlinear 
interactions that follow the initial instability. Previous research on 
ripples has generally relied on highly simplified continuum models or on 
stochastic or molecular dynamics simulation. By means of a new deterministic 
continuous model that seems to better describe the essential physics, we here 
investigate salient nonlinear properties of ripple formation. 

As is well known, Aeolian ripples are oriented perpendicularly to the wind
direction. Mature ripples are asymmetrical in cross section: their stoss
(upwind) slopes are typically much less steep 
than the
shorter lee (downwind) slopes \cite{And90,Sharp,WernHLA}. 
The  steepness of the lee slopes cannot exceed and usually
does
not reach the sand angle of repose.
The ripples have convex stoss slopes, concave lee slopes, and
flattened crests which usually end with a brink.  
  Since smaller ripples of the same shape have smaller volume to surface
ratios, they are translated faster by the wind and can overtake the larger 
ripples.
A possible merger results in gradual elimination of small ripples and
in growth of ripple wavelength. 

To analyze the mechanics of sand transport, which occurs whenever the wind
is sufficiently strong, it is convenient to distinguish two 
types of sand grain movement: saltation and reptation (or creep)
\cite{Bagnold,And90}. 
Saltating grains move by long trajectories that end in high-energy
impacts with the surface.  These impacts take place at almost uniform shallow
angles of descent varying from 9 to 15$^{\circ}$ to the horizontal
depending on the 
wind strength and grain size \cite{WR}. After an impact, a saltating grain  
usually rebounds sufficiently
high to be accelerated by the wind again and continues its saltation.  These
grains gain energy from the wind and transfer part of it to the sand bed
on impact. Each impact may cause ejection of one or several low energy
(reptating) particles from the bed surface. Reptating particles make a short
hop, usually jumping or rolling for several millimeters or less, and stop. 
The exchange flux between saltating and reptating grain populations is 
supposed to be small \cite{And87,LW}.

According to the hypothesis put forward by Bagnold \cite{Bagnold}, the ripple
wavelength is equal to the mean length of saltation jump. However, this 
claim has been challenged by numerous researchers (see, e.g., 
\cite{And90,And87}) and it is now commonly accepted 
that the essential physics lies in the variation of reptation flux.
The role of saltation, whose trajectories are many times longer than the ripple
wavelength, is indirect. The oblique, almost unidirectional bombardment by
saltating particles supplies the energy necessary for reptation. 
Wind strength determines the intensity of saltation; the 
probability of direct entrainment of particles into reptation by 
wind is small \cite{Bagnold}. 

A model, based on these views, was proposed by Anderson
\cite{And87}. Linear stability analysis of this model showed that the
initial ripple wavelength is determined by, and is several times larger than,
the mean length of reptation. Unfortunately, the model yields unrealistic
results at the nonlinear stage of ripple growth, which begins very early.
It has been suggested \cite{And90} that the model can be improved by allowing the 
reptating grains to
continue rolling upon the bed surface after landing, and not to stop immediately 
as was assumed originally. Although no
such continuous model has been developed, employing a similar approach in
molecular dynamics computer simulations of sand ripples \cite{LW} was quite
successful. 

Interesting results on modeling different aspects of sand ripples dynamics
have been obtained by means of cellular automata models or molecular dynamics
simulations \cite{LW,ForrestH,AB,WG}. It was even claimed
\cite{WG} that no continuum mechanics or deterministic model can capture
the main feature of ripple self-organization: the increase of scale in
time due to merging of ripples.  Below, we show that this general conclusion,
based on a schematic discrete model of ripple merging, is wrong.  The
continuous deterministic model proposed herein provides a better
description of the physical process than the simplified 
stochastic model \cite{WG}. Our model reproduces the 
asymmetrical ripple shape and is able to simulate not only merging of
ripples but also a more complicated, soliton-like mode of ripple
interaction which can be equally significant for ripple
self-organization. A somewhat similar behavior has been 
observed in molecular dynamics simulation of sand ripples \cite{LW}. 

Let us mention here also the analytical model of sand ripples \cite{HW}. In
this model, the upwind slope of a ripple is given by a smooth solution of the
diffusion equation with a negative diffusion coefficient. Such a solution is,
however, unstable.  Furthermore, in the absence of wind the ripple shape in 
\cite{HW} is
described by a well-posed diffusion equation. This is also unphysical: 
according to this model 
ripples diffuse and disappear, since surface particles continue to roll down the
slopes. Real sand ripples are, of course, metastable. 

To derive a model allowing for metastability, it is necessary to take into
account that the surface flux of granular material is not determined solely by 
the local surface slope. Models involving an additional variable,
surface flux or the density of rolling particles, have recently been derived to
simulate quasistationary evolution of a growing pile shape \cite{Pr} or to
model the dynamics of pile surface in more detail and on a shorter
spatio-temporal scale (\cite{BCPE}, see also \cite{Mehta}).  In our model of
Aeolian ripples  we use a similar approach to describe the flow of
rolling particles. 

\section{Mathematical model}
In accordance with the physical picture of sand transport described above,
we assume that there exists a uniform flux of saltating grains which
reach the bed surface $y=h({\bf x},t)$ at a low angle $\gamma$ to horizontal. 
 The impacts cause erosion of this surface.
The erosion rate $f$ is proportional to the 
impact
intensity, which depends on the surface orientation with respect to
the direction of saltation. Let the saltating particles strike an 
inclined surface at an angle $\theta$. Then $f$ is 
proportional to $\sin
\theta$ and we can write 
$$
f=f_0\frac{\sin\theta}{\sin\gamma},
$$
where $f_0$ is the rate of erosion of a horizontal surface 
(determined by the intensity of saltation). However, if
the surface has sufficiently steep slopes, some parts of it may be in
shadow and unreachable by saltating grains. In this case we set 
$f=0$; shadowing introduces non-locality into this problem. 

A reptating particle, ejected by an impact at a point ${\bf x}$, makes a
jump and lands on the bed surface at a point ${\bf y}$ with probability
density $p$ given by the "splash function", $p=p_{\alpha}({\bf x},{\bf y})$, first
introduced in \cite{UH}. This function, which will be specified below,
depends, in particular, on the surface slope at the point of impact. 
In our model, the splash function also accounts for all
the anisotropy induced by a chosen wind (saltation) direction. 

Upon landing, reptating particles do not stop immediately  
but may roll away, although usually not far from the landing point.
Let $R({\bf x},t)$ be the effective surface density of rolling particles
($R\,d{\bf x}$ is the volume which 
particles, presently rolling over the part of the free surface above the 
area $d{\bf x}$, 
would occupy in the sand bed). When they stop, the rolling particles
are incorporated into the motionless bed. Following \cite{BCPE}, we 
denote by $\Gamma[h,R]$ the rate of rolling-to-steady state transition and 
write 
the mass conservation equations for the sand bed and for the population of 
rolling particles:
\beqa
\partial_th=\Gamma[h,R]-f,\label{heq}\\
\partial_tR+\nabla \cdot {\bf J}=Q-\Gamma[h,R].\label{nonstR}
\eeqa
Here ${\bf J}$ is the horizontal projection of the flux of
rolling particles, and the source term 
\beq
Q({\bf x},t)=\int 
f({\bf y},t)p_{\alpha}({\bf y},{\bf x})\,d{\bf y}
\label{Q}\eeq 
gives the intensity of "rain" of falling reptating particles. 

We assume 
that reptating particles lose most of their momentum in collision with the
rough bed surface. Neglecting inertia, we postulate that upon landing the
particles roll in direction of the steepest descent and that the steeper the
slope the faster they roll.  
 The simplest form \cite{J} of the flux ${\bf J}$ is,
therefore, 
$${\bf J}=-\mu_0R\nabla h,$$ 
where $\mu_0$ is a constant "mobility"
of particles. 

The rate of rolling-to-steady state transition, $\Gamma$, depends on
stability of a particle on inclined sand bed surface and on the amount of
rolling particles. Rolling particles never form a thick layer on the surface
during the ripple growth: there is only a small amount of them at each time
moment. It can be assumed that, for a fixed free surface incline, the rate of
rolling-to-steady state transition is proportional to the amount of rolling
particles on the surface, R. Since the exchange rate cannot depend on the
free surface slope orientation, we further assume $\Gamma$ is a (smooth)
function of $|\nabla\, h|^2$.  The steeper the free surface is, the easier do
particles roll down and, correspondingly, the lower is the rate of 
rolling-to-steady 
state transition. For slopes steeper than the sand angle of repose,
$\alpha_r\approx 30^{\circ}$, rolling sand grains do not stop at all. Taking
these arguments into account we assume, as a simple but physically reasonable
approximation, that $\Gamma$ is also proportional to
$\tan^2\alpha_r-|\nabla\, h|^2$ and obtain
 $$
\Gamma=\kappa_0R\left(1-\frac{|\nabla\, h|^2}{\tan^2\alpha_r}\right),
$$
where $\kappa_0$ 
characterizes particle stability upon a horizontal surface.

We now rescale the variables,
$$t'=f_0t,\ \ \ R'=\frac{\kappa_0}{f_0}R,\ \ \ f'=\frac{1}{f_0}f,\ \ \
{\bf J}'=\frac{1}{f_0}{\bf J}, \ \ \ {\Gamma}'=\frac{1}{f_0}\Gamma ,
$$
and obtain
\beq
{\bf J}=-\nu R\nabla h,\ \ \ 
\Gamma=R\left(1-\frac{|\nabla\, h|^2}{\tan^2\alpha_r}\right),
\label{JG}\eeq
and $f=\sin \theta/\sin \gamma$, or $f=0$ in a shadow. Here 
$\theta=\theta({\bf x},t)$ is the 
angle at which the saltating particles strike 
the bed, and $\nu=\mu_0/\kappa_0$ is a constitutive dimensionless 
parameter characterizing the competition 
between mobility and stability of dislodged grains on the sand 
surface.
The rescaling leaves the equation (\ref{heq}) invariant while the 
equation (\ref{nonstR}) takes the form
\beq
\frac{f_0}{\kappa_0}\partial_tR+\nabla \cdot {\bf J}=Q-\Gamma[h,R].
\label{fk}
\eeq
We suppose that the rate of sand surface erosion caused by 
saltation, $f_0$, is usually much 
smaller than $\kappa_0$, the coefficient determining the rate at which 
rolling particles come to a stop on the 
horizontal 
surface and are absorbed by the motionless bulk (that is why the layer of rolling  
particles is so thin).
Neglecting the small term in equation (\ref{fk}) we arrive at a 
quasistationary mass balance equation for rolling particles,
\beq \nabla \cdot {\bf J}=Q-\Gamma[h,R]. \label{Req} \eeq

\section{Splash function}
To complete the model (\ref{heq}), (\ref{Q}), (\ref{JG}), and 
(\ref{Req}) we need to specify the splash function.
Although not much is known about this function, previous studies
(see, e.g., \cite{And87,LW,ForrestH}) suggest that the system is not very 
sensitive to the details of splash function behavior and that an approximation,
sufficient at least for qualitative simulation, may be
obtained by combining the existing experimental data and simple physical
arguments. We limit our consideration to the one-dimensional (1D) case.

Collision of quartz grains with a sand bed has been studied experimentally by
Willetts and Rice \cite{WR}. It was found that ejection of reptating grains
from the bed depends only slightly on the incident angle of attack, which varied
in these experiments in accordance with the systematic changes of the
bed inclination. (The angle of descent of saltating particles to the horizontal
was constant). For various incident angles,
ejection occurred at approximately the same mean angle to the bed surface,  
$m_{\phi}\approx 50^{\circ}$, 
with standard deviation $\sigma_{\phi}\approx 40^{\circ}$, and the same mean 
velocity of ejecta. 

We use these results to crudely reconstruct the dependence of the splash
function on the bed surface inclination in 1D case. First, we define (somewhat
arbitrarily, see \cite{Arbitrary}) the density of ejection angle distribution,
$p=p(\phi)$, providing for the mean and standard deviation values 
as found in \cite{WR} (see  Fig. 1a). For simplicity, we further assume 
that  particles are ejected from the 
bed with the same initial velocity $v_0$ at different angles $\phi$,
that they then follow simple ballistic trajectories 
$x=x_0+v_0\cos(\alpha+\phi)t,\ 
y=h(x_0)+v_0\sin(\alpha+\phi)t-\frac{1}{2}gt^2,$
 and hit the  
surface at a horizontal distance $s$ from the ejection point.
Here $\alpha=\tan^{-1}(\partial_xh)$ is the surface angle at the ejection 
point $x_0$, $g$ -- acceleration of gravity. 
Assuming the surface curvature is small,  $|\partial^2_{xx}h|\ll g/v_0^2$, 
we approximate
$h(x_0+s)$ by $h(x_0)+s\partial_xh(x_0)$ and find the jump length
$$s=\frac{v_0^2}{g}(\sin 2\{\alpha+\phi\}-2\cos^2\{\alpha+\phi\}\tan\alpha).$$
Making use of the probability density $p(\phi)$, we can now 
calculate numerically
the mean and the standard deviation of reptation jump length, $m_r(\alpha)$ and 
$\sigma_r(\alpha)$, for any bed surface inclination 
$\alpha$, up to the value of a proportionality coefficient $v_0^2/g$. This 
factor is eliminated from the final dimensionless formulation of the model
by choosing the unit of length equal to the mean reptation length at the 
horizontal bed surface. Thus $m_r(0)=1$ by definition. The functions 
$m_r(\alpha)$ and $\sigma_r(\alpha)$ are shown in Fig. 1b.
Finally, we approximate the splash function
$p_{\alpha}$ by the density of a corresponding normal distribution,
$$
p_{\alpha}(x_0,y)=\frac{1}{\sigma\sqrt{2\pi}}\exp\left(-\frac{1}{2}
\left[\frac{y-x_0-m}{\sigma}\right]^2\right),
$$
where $\sigma=\sigma_r(\alpha)$, 
$m=m_r(\alpha)$, and $\alpha=\tan^{-1}\partial_xh(x_0,t)$.

\section{Linear stability analysis}
To analyze the initial instability of a horizontal sand bed,
we assume that $h$ and its derivatives are small
and linearize the model. Up to the second order terms,
$\Gamma[h,R]=R$ and $f=1+k_{\gamma}\partial_xh$, where 
$k_{\gamma}=\cot{\gamma}$,
so Eq. (\ref{heq}) yields 
$R=1+\partial_th+k_{\gamma}\partial_xh.$

For small surface slopes, the standard deviation of reptation length does 
not change much (see Fig. 1b) and we set 
$\sigma_r\equiv \sigma_r(0)=1.25$. Knowing the dependence $m_r(\alpha)$
(Fig. 1b), it is easy to find numerically that for small slopes $m_r\approx 
1-k_m\partial_xh$, where $k_m=2.01$. 
 
Let $p_0(x)$ be the density of the normal distribution
$N(1,\sigma^2_r(0))$. It is not difficult to show that
$p_{\alpha}(x,y)=p_0(y-x)+k_mp'_0(y-x)\partial_xh(x)$ plus the higher order terms
($"'"$ means derivative).
The linearized  Eq. (\ref{Q}) takes the form
$Q=1+k_{\gamma}p_0*\partial_xh+k_mp'_0*\partial_xh,$
where "*" is the operator of convolution.

Substituting the linear approximations for $\Gamma,\ R,$ and $Q$ into Eq. 
(\ref{Req}) we obtain, up to the second order terms,
$$\partial_th=\nu\partial^2_{xx}h 
+k_\gamma(p_0*\partial_xh-\partial_xh)+k_mp'_0*\partial_xh $$
We can now apply the Fourier transform  and
substitute $h=e^{\lambda t +i\omega x}$ to find the dispersion 
relation
$$
\lambda(\omega)=-\nu\omega^2+k_{\gamma}iw(\widetilde{p}_0-1)-
k_m\omega^2\widetilde{p}_0,    
$$
where $\widetilde{p}_0=exp(-[\omega\sigma_r(0)]^2/2-iw)$ is the Fourier 
transform of $p_0$. Note that with $\nu=k_m=0$ one gets the 
dispersion relation for Anderson's model \cite{And87}.
The initial ripple wavelength can be calculated as  $l_0=2\pi/\omega_0$, where 
$\omega_0$ is the wave number at which the expression
$$Re 
\lambda(\omega)=-\nu\omega^2+\omega(k_{\gamma}\sin \omega-k_m\omega\cos \omega)
e^{-{(\omega\sigma_r(0))^2}/{2}}$$ 
attains a positive maximum. This maximum exists and the flat surface is 
unstable if \beq
k_{\gamma}>k_m+\nu .\label{ineq}
\eeq
To explain this result we note that ripples grow because of the geometrical 
effect of greater impact and
ejection flux on upwind-oriented slopes than on downwind-oriented slopes
\cite{ForrestH}. This non-uniformity increases if the saltation angle of
attack becomes smaller ($k_{\gamma}$ increases). On the other hand, the greater
$\nu$ and $k_m$ are, the more significant are, respectively, smoothing effects 
due to rolling of 
dislodged particles down the surface slopes and scattering of ejected particle
trajectories by an uneven sand surface. 
Condition (\ref{ineq}) indicates when 
the instability 
prevails. 

As it was mentioned above, the saltating particles usually hit 
the sand surface at almost uniform angle to the horizontal, varying from 9 to
15$^{\circ}$. In the examples below we use the mean reported value of this
angle, $\gamma=12^{\circ}$. Using the stability condition (\ref{ineq}), 
it is easy to calculate that the flat surface is unstable if $\nu < 2.69$.

\section{Nonlinear dynamics}
To study the ripple evolution further we solved the nonlinear system 
(\ref{heq}), \ref{Q}), (\ref{JG}), (\ref{Req}) numerically, 
assuming periodic boundary conditions
and using an implicit  finite-difference approximation. 
The initial evolution of a slightly disturbed flat surface 
obeys the linear theory: after a short initial stage the fastest growing mode,
having the wavelength predicted by the linear stability analysis,
dominates. 

The growing ripples remain almost symmetric until, shortly before 
the 
appearance of first shadow regions, the downwind slopes become steeper.
This asymmetry develops quickly as the ripples continue to grow. Further
evolution is accompanied by the increase of the ripple wavelength (see
Fig. 2). It can be seen that the downwind translation of ripples gradually
slows down as their size increases. Although our model is much simplified 
in many respects, the calculated mature ripple shape is
similar to that of the real sand ripples \cite{Sharp}: for $\nu=2$ we 
obtained convex
stoss slopes inclined at about $11-13^{\circ}$, flattened crests, and
slightly concave lee slopes with the mean maximal inclination
$26-27^{\circ}$; the ripple index (length to height ratio) was about
14. For $\nu=1$ the results are qualitatively similar, although the mean
maximal inclinations are 15$^{\circ}$ and 28-29$^{\circ}$  
for the
stoss and lee slopes, respectively, and the ripple index is
smaller. 

The most interesting part of ripple dynamics is the
mechanism of ripple merging and self-organization. The simulations show
that simple merger takes place only if the overtaking ripple is much
smaller than the bigger "overtaken" ripple, which moves more slowly. Otherwise 
another, 
more complicated scenario is usually realized.  As a smaller ripple reaches the
larger, the trough between them becomes shallow and a "two-headed" long
ripple appears. Then the "downwind head", which originally formed the
larger ripple crest, starts to move forward as a separate small ripple and
runs ahead.  Two new ripples emerge from this recombination. The ripple
that is left behind is larger than the larger of the two ripples before
merger. The ripple that runs away is smaller than the smaller one before
this event (Fig. 3).  Sometimes the trough between the two merging ripples
disappears before the generated long ripple becomes unstable. However, soon
there appears a new trough near the end of this ripple crest and a
small running away ripple develops. Since it is smaller, this ripple
proceeds even faster and soon meets another large ripple on its way.
Complete merger is now more probable, since the overtaking ripple became
smaller.  However, another recombination may yet occur before the material
redistribution between ripples is completed. Obviously, such a mechanism of
ripple interaction produces a regular ripple array structure more 
efficiently than a simple merger of ripples.
 
In our opinion, this scheme of ripple interaction gives also a likely
explanation to the appearance of small secondary ripples in wind tunnel
experiment \cite{SL}. Indeed, the appearance of such ripples due to the
backward eddy flow behind a ripple, as is suggested in \cite{SL}, seems
hardly possible. Sand ripples are so shallow that most probably there is no
backward flows in their shadows.  Even if small backward eddies existed,
they could never cause saltation of sand grains against the main wind
direction, and thus could not produce any sand ripples.  As follows from
our simulation, no backward flows are necessary: small ripples appear as a
means of redistribution of mass during the ripple array reconstruction
leading to the wavelength growth. 

 \section*{Acknowledgments}
The author appreciates discussions with Yu. Shtemler and I. Rubinstein.
This work was supported by the Blaustein International Center for Desert Studies.

\begin{figure}
\caption{{\it a} -- density of ejection angle distribution, $p(\phi)$;
{\it b} -- the mean (solid line) 
and standard deviation (dashed line) of reptation jump
length as functions of surface inclination.}
\end{figure}

\begin{figure}
\caption{Flat surface instability and formation of sand ripples.
Wind direction is from left to right; the unit of length is 
the mean length of the reptation jump; $\nu=2$.} \end{figure}

\begin{figure}
\caption{A typical ripple interaction; see 
the region bounded by the thin line.
To show the details the ripples are stretched in the vertical direction.}
 \end{figure}


\begin{thebibliography}{99}
\bibitem{J} The diffusion component of surface flux, $-D\nabla 
R$, is omitted. We found that if such a term is made significant,
the model yields a somewhat distorted ripple shape.
\bibitem{Arbitrary} This function has been identified 
by the known mean and variance values from a two-parametric 
family of probability densities,
$p(\phi)=C\exp(-\frac{1}{2}\left\{\frac{\phi-a}{b}\right\}^2),$
where $C$ is the normalizing coefficient and $0\le \phi\le 180^{\circ}$. The 
identified
values of parameters are: $a=-40^{\circ}, b=80^{\circ}$.
\end{thebibliography}
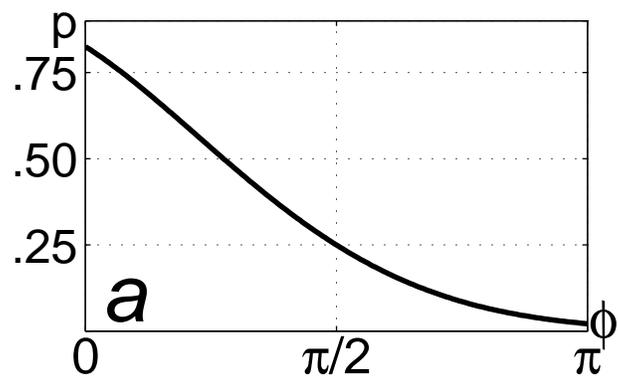
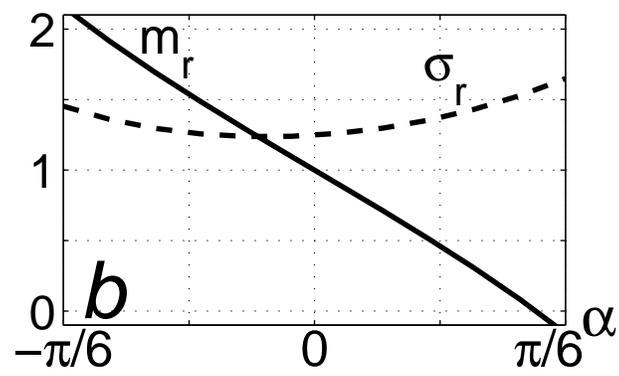
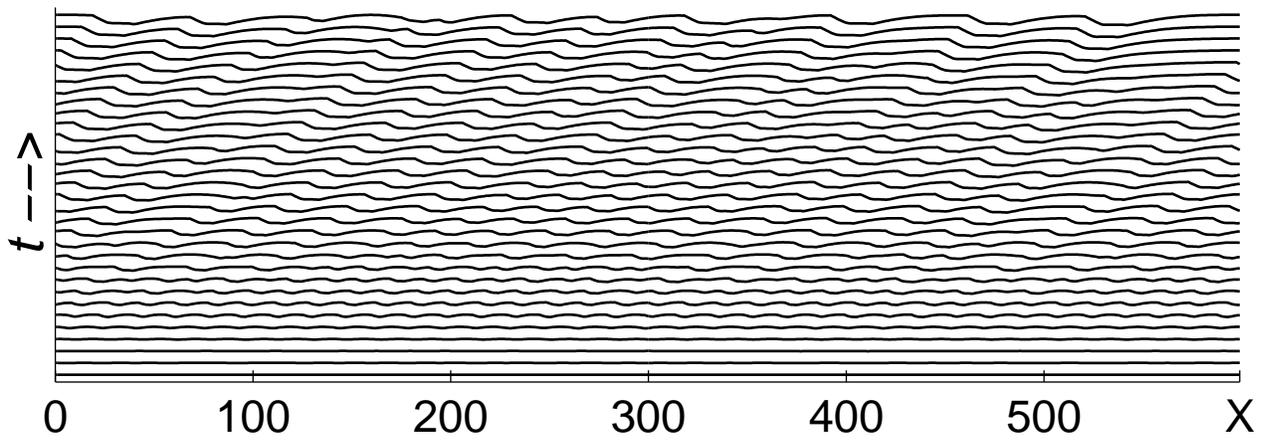
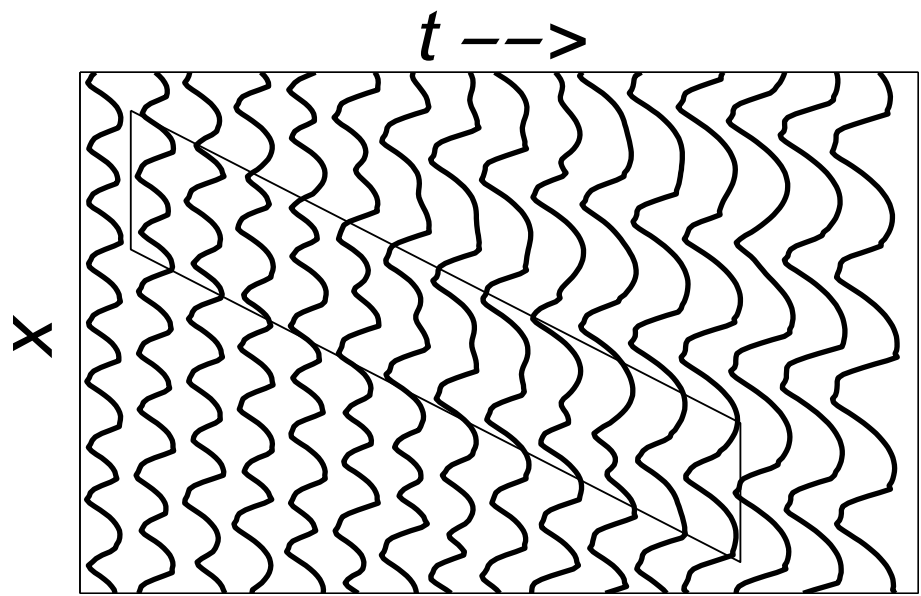
\end{document}